# From Global Value Chains to Local Jobs: Exploring FDI-induced Job Creation in EU-27

**Magdalena Olczyk**
Gdańsk University of Technology, Poland
Magdalena.Olczyk@pg.edu.pl

**Marjan Petreski**
University American College Skopje, North Macedonia
Gdańsk University of Technology, Poland
Finance Think – Economic Research & Policy Institute, Skopje, North Macedonia
marjan.petreski@uacs.edu.mk

## Abstract

This study explores the differential impacts of global value chain (GVC) participation on foreign direct investment (FDI)-related job creation in EU-27, emphasizing the role of sector-specific and regional factors. The study is based on a rich set of project-level data on FDI-generated jobs. It utilizes a labor demand function estimated through GMM estimator to account for endogeneity. Results indicate that forward GVC participation significantly boosts FDI-related job creation by enhancing domestic value-added and production capacity. However, this effect is moderated by sector-specific characteristics such as productivity or wages. Conversely, backward GVC participation, characterized by reliance on foreign inputs, generally reduces FDI-generated jobs due to lower domestic labor requirements and diminished competitiveness. Despite this, the negative impact of backward GVC participation on employment becomes less significant when regional diversification is considered, highlighting the importance of regional factors like infrastructure and skilled labor. The study also finds that the impact of GVC participation on employment varies with EU membership status and sectoral characteristics, with old EU member states and high-tech sectors benefiting more from forward GVC integration. In contrast, new EU member states and low-tech sectors face greater challenges, particularly with backward GVC participation.

**Acknowledgement:** Declaration of generative AI and AI-assisted technologies in the writing process: During the preparation of this work the author(s) used ChatGPT in order to proofread the text. After using this tool/service, the author(s) reviewed and edited the content as needed and take(s) full responsibility for the content of the publication.

1. Introduction

Modern globalization is closely linked to global value chains (GVCs) and even after the Great Recession and the COVID-19 pandemic, GVCs still remain an essential feature of the global economy (Chepeliev et al. 2022). The development of GVCs not only influences product and service markets, but also directly and indirectly affects labor markets (Farole et al. 2018). It is why the extensive economic literature dealing with the impact of trade on labor market outcomes has undergone a remarkable development. It evolved from country-specific frameworks to more complex industry- and firm-level models that take into account the characteristics of firms' product and labor markets, but also focuses on new theoretical frameworks that address global value chains and their effects on labor markets (Aleman-Castilla 2020).

The recent literature on GVCs and the labor market focuses on the relationship between countries' integration into GVCs and labor market outcomes. Most (but not all) studies emphasize the benefits of a country's participation in GVCs for wages, employment, working conditions, labor productivity, and labor informality, especially in the case of developing economies (Shepherd, 2013; Kummritz and Quast 2016; Hollweg, 2019; Guschanski and Onaran 2023; Mingyang et al. 2023). Several factors determine the way in which a country's labor market is affected by GVC integration, including the country's level of development, the type of sector, the domestic skills base and the institutional environment (Farole et al. 2018; Carneiro et al. 2024). Extensive empirical evidence suggests that the negative effects of GVC participation on wages mainly affect low and medium-skilled workers, as higher participation is associated with a higher intensity of routine tasks in offshorable occupations, which in turn is strongly associated with lower wages (Lewandowki et al. 2023). GVC participation and upstream specialization are associated with higher wages, but only in developed countries (Wolszczak et al. 2023). In addition, participation in GVCs provides more job opportunities for rural, female and low-skilled workers (Korwatanasakul et al. 2020; Cao et al. 2022) and supports the inclusion of some groups of workers in the formal labor market (McCaig and Pavcnik, 2018) and promotes gender equality (Bamber and Staritz, 2016; Nur amd Martínez-Zarzoso, 2022).

Despite the large number of studies on this topic, there is no consensus in the literature on the direction and extent of the impact of GVCs on employment (Carneiro et al. 2024). There are several reasons for this. On one hand, it is inherently difficult to isolate the net effects of GVCs from the role of trade policy, changes in consumer demand or technological innovation (Escaith et al. 2018). On the other, the multiplicity of contradictory effects arising through different channels creates additional difficulties, especially when outcomes manifest in opposite directions (Farole et al. 2018). Finally, Shingal (2015) points out that GVCs and employment are a complex issue because while it is influenced by many factors, the labor market losses of GVC integration are visible and concentrated, the gains are more hidden and untapped.

Despite the difficulties mentioned above, this article aims to assess the impact of integration into the GVCs on employment, i.e. on domestic job creation in the EU. We focus exclusively on the creation of new jobs resulting from foreign direct investment (FDI). There is only one study in the literature that looks at the impact of GVC integration on job creation through FDI (Bail et al. 2024), but the authors used different approach: a cross-country input-output model. Hence, this paper enriches the existing literature in two ways.

First, previous studies of the effects of countries' integration on employment have always been based on data on total domestic employment. The weakness of this approach is that employment growth can result from many factors unrelated to participation in GVCs, such as micro factors: the age of workers, the level of education, propensity to invest in human capital, the size of a firm; as well as macro factors: economic conditions, employment policies, institutional changes in the labor market. In our study, we use data on job creation through the inflow of FDI, which by its nature is closely linked to participation in trade and GVCs (WTO, 2014).

Second, existing studies that consider country heterogeneity compare the effects of GVCs integration on employment in developing countries with those in developed countries. Our analysis refers to the EU-27 members that belong to the developed countries according to the income criterion of the World Bank or the HDI criterion. However, Kordalska et al. (2022) find a dichotomy between the EU countries according to their GVC participants: the Central and Eastern European countries are predominantly specialized in fabrication tasks and serve as "factory economies", while the EU-15 countries are mainly engaged in knowledge-intensive activities in pre- or post-fabrication – a characteristic of "headquarters economies". In our study, we want to take into account the 'functional' heterogeneity of EU-27 (old vs new EU members) and examine whether it is relevant for the impact of countries' integration into GVCs on employment.

Finally, we use a new approach in our study based on a new micro-dataset from the cross-border investment monitor FDI Markets database maintained by the Financial Times. The database captures individual greenfield FDI projects from 2003 to 2023.

The paper is organized as follows: The next section presents a literature review on relationship between GVCs integration and employment. Section 3 details the data and the empirical model. Section 4 contains the results and offers a discussion. The final section concludes and offers some policy implications.

## 2. Literature review

The theoretical background of our paper is provided by the theories on trade in intermediate goods (Helpman, 1985), fragmentation of production (Jones and Kierzkowski, 2001) and the recent studies on the determinants of GVC participation (Gereffi and Fernandez-Stark, 2018). They all emphasize a crucial role of financial issues (access to finance) / investments in trade in intermediate goods. Some authors assign a special role to FDI as the main driver of GVC expansion in the last decade (Baldwin, 2011; WTO, 2014). Also, numerous empirical studies confirm that FDI has a positive impact on GVC participation (Li, 2019, for 63 countries (2005–2016); Efogo et al. 2022, for 43 developing countries (2010–2019)). According to Qiang et al. (2021), the importance of countries in the global FDI network correlates strongly with their importance in the GVC network. Although countries follow different development paths, their growing importance in the GVC network is often preceded by an increasing integration of FDI with the rest of the world. Additionally, Adarov (2021) finds that both GVC and FDI networks have a fairly similar core-periphery structure, with a relatively small number of highly influential countries forming the highly interconnected core.

Research on the role of multinational enterprises (MNEs) in GVCs also provides a justification for using employment data, i.e. new jobs created by FDI in our research.

OECD et al. (2014) point out that MNEs are an important driver of the GVC growth, i.e. in their quest to expand operations, MNEs engage in FDI and establish subsidiaries in different countries. MNEs evolved beyond the horizontal and vertical dichotomy and became a hybrid version of both categories, albeit at the same time they increasingly function as networks within the international production networks of GVCs (Cadestin et al. 2018). Studies analyzing the role of MNEs in GVCs, compared to domestic firms, in some European countries (like Germany or UK) revealed a dominant presence of MNEs in the provision of value added in the manufacturing and service sectors through complex GVC networks (Gao et al. 2023). In addition, Fortanier et al. (2020) found higher import content of exports in MNEs compared to domestic companies, what can be the sign of the MNEs stronger creation of local backward linkages. Finally, based on a cross-country input-output model, Bail et al. (2024) conclude that MNEs make a significant contribution to employment in GVCs, amounting to 7.4–8.0%, especially in high-income economies (13.8%) and high-tech manufacturing (22.6%).

To explain the relationship between GVC participation and employment driven by FDI, we refer to the research related to the impact of the former on the latter. In the theoretical literature, the roots for the impact of GVC on the level and structure of employment are linked to classical trade theories. The Hecksher-Ohlin theory states that the comparative advantage of countries results from the relative abundance of factors and this comparative advantage is influenced by the interaction among countries' resources (the relative abundance of factors of production). Melitz model (2003) shows that a free trade leads to a reallocation of resources in favor of the export sector, where skilled labor predominates, resulting in higher demand for this type of labor and higher wage levels. Also, Bernard et al. (2007), who joined the Helpman and Krugman's (1985) trade model with the concept of heterogeneity of firms, showed that trade leads to increased demand for labor in sectors with comparative advantage, in contrast to job destruction in sectors without a comparative advantage. Grossman and Rossi-Hansberg (2008) identify three simultaneous effects of interrelation between offshoring and the labor market, namely: the productivity effect, the relative price effect (similar to the Stolper-Samuelson theorem) and the labor supply effect (which arises from the lower demand for domestic labor as a result of the internationalization of production).

It should be remembered that trade theories are based on many strong and often unrealistic assumptions and largely refer to trade in final goods and not to intermediate goods, which are the main objects of trade in GVCs. Taglioni and Winkler (2016) try to take into account the specificity of GVC and restore the assumptions of comparative advantage theory in times of fragmented production, and identify three channels for the potential impact of GVC on labor markets in developing countries: the demand effect strictly related to employment (MNCs participating in GVC increase the demand for skilled workers in the local labor market), the skill-upgrading effect (local workers receive training and upskilling from MNCs) and the spillover effect (local workers move from MNCs to local companies and bring the acquired skills and knowledge with them).

Farole et al. (2018), in turn, characterize three different effects of GVC participation on the total labor in advanced economies, which often have opposite effects on the labor market: the substitution effect: offshoring shifts part of production activities abroad and replaces domestic with foreign labor, reducing domestic labor demand; the productivity effect: outsourcing enables a higher degree of specialization and improves labor productivity, reducing the demand for labor per unit of output; and the scale effect:

offshoring lowers production costs, leading to lower prices and higher demand, which in turn increases the demand for labor to produce higher output. According to Jones at al. (2019), it is difficult to assess the net impact of GVCs on labor demand due to the variety of factors which simultaneously influence labor market.

Most recent research focuses on the identification of factors that can trigger or limit the impact of GVC participation on employment and the effects of integration in GVC on domestic jobs, considering the countries' and sectors' heterogeneity (Meng et al. 2020; Pahl et al. 2022). Analyses examining the nature of a country's integration into GVCs—whether through forward participation as an exporter of intermediate goods for production in GVCs or backward participation as an importer of production inputs for creating final goods—along with the chain's activity and relative position, and their impact on domestic employment, have increasingly attracted researchers' attention (Oladapo and Raifu, 2022; Mingyang et al. 2023).

The empirical literature offers numerous analyses on the impact of GVC participation on employment, which can be categorized into three main areas:

1. Individual country studies, such as Banga (2017) for India, Dine (2019) for Turkey, Long et al. (2019) for Vietnam, Korwatanasakul et al. (2020) for Thailand, Oladapo and Raifu (2022) for Nigeria, Guha-Khasnobise et al. (2023) for India, Ndubuisi and Owusu (2023) for South Africa, and Pan (2020) for the U.S.

2. Group of countries: Research covering multiple nations includes Shepherd and Stone (2013) for OECD and emerging economies, Jones et al. (2019) for developed countries, Shepherd (2013) and Hollweg (2019) for developing countries, Grunella et al. (2017) for Euro area countries, Paul et al. (2019) for four African countries, Nikulin and Szymczak (2020) for 10 Central and Eastern European countries, Camarero et al. (2022) for EU countries, Mingyang et al. (2023) for 42 countries, and Duarte and Castilho (2024) as well as Szymczak and Wolszczak-Derlacz (2021) for 46 countries with different income levels. Winkler et al. (2023) focus on low- and middle-income countries.

3. Sector-specific studies: Shepherd (2013) examines the electronics sector in Asia and services in Chile, Shingal (2015) explores the garment and textile industries in Vietnam, Bangladesh, and South Africa, as well as horticulture in Kenya and South Africa. Lopez-Acevedo and Robertson (2016) focus on the apparel sector in South Asia.

It is not easy to find a clear answer as to how GVC participation influences employment. Feenstra and Hanson (1996) show that outsourcing leads to an increase in the relative demand for skilled labor in both developed and developing countries in general. Carabello and Jiang (2016) also discover that the jobs embodied in exports are moving away from those with low-skilled labor towards those with high- and medium-skilled labor. The skill-orientated nature of GVC trade is accompanied by an increasing complexity of global supply chains and an increased use of skill-intensive inputs.

The literature consistently highlights significant positive effects of GVC activities on employment across various contexts. For instance, Grunella et al. (2019) for Eurozone countries, Jones et al. (2019) for developed countries, Pan (2020) for the U.S., and Szymczak & Wolszczak-Derlacz (2021) for high-income nations all report such findings.

In developed countries, GVC growth benefits primarily highly skilled labor and capital owners (Grunella et al. 2019; Carpa & Martínez-Zarzoso, 2022).

For developing nations, the evidence is largely positive. Shepherd (2013) finds that GVC participation supports labor markets, while Hollweg (2019) and Mingyang et al. (2023) confirm its role in job creation, not only in export sectors but also through linkages between exporters and domestic input providers. However, Farole (2016) presents a more nuanced view, suggesting mixed effects, with both winners and losers. Duarte and Castilho (2024) further show that the impact of GVCs is country-specific and depends on whether a nation is engaged through backward or forward linkages.

There is no clear consensus on which group of countries benefits most. Mingyang et al. (2023) argue that GVCs have a stronger employment impact in developing nations, while Carneiro et al. (2024) suggest otherwise, pointing to limitations in developing economies, such as the prevalence of low-skilled labor and backward integration, which can substitute domestic production, negatively affecting employment.

The nature of GVC involvement—whether forward or backward—also shapes labor outcomes. Forward participation, driven by medium- and high-skilled labor, tends to negatively impact low-skilled employment while boosting high-skilled jobs (Farole et al. 2018). In contrast, backward participation, often supported by low-skilled labor and foreign capital, shows mixed results. For developed economies, backward GVC participation correlates inversely with employment (Farole 2016), while in developing countries, such as those studied by Oladapo and Raifu (2022), it may positively impact job growth.

Sectoral differences also play a key role. Winkler et al. (2023) emphasize manufacturing's central role in employment creation, particularly in emerging economies, while business services exports are more relevant for high-income nations. Mingyang et al. (2023) further note that employment effects are industry-specific, with manufacturing showing stronger positive impacts compared to services, where backward integration often results in fewer jobs. Finally, Duarte and Castilho (2024) conclude that the GVC-employment relationship is most evident in manufacturing, where jobs are created at the early stages of production chains.

## 3. Empirical model and data

### 3.1. Empirical model

To understand if GVC participation has a role to play for the job creation by foreign direct investment (FDI), we employ the following labor demand function:

$$\ln(Jobs_{sit}) = \beta_1 GVC_{sit} + \beta_2 GVC\_pos_{sit} + \beta_3 X'_{it} + \beta_4 Z'_{sit} + \alpha_s + \alpha_i + \alpha_t + \varepsilon_{sit} \quad (1)$$

Whereby: $\ln(Jobs_{sit})$ is the natural logarithm of the number of jobs created in sector $s$, country $i$, and year $t$ by FDIs, the sector being at the two-digit NACE Rev. 2 level. $GVC_{sit}$ is the GVC participation index and $GVC\_pos_{sit}$ the position of GVC participation (e.g., upstream or downstream position). $X'_{it}$ is a vector of control variables specific to country $i$, and year $t$, while $Z'_{sit}$ includes sector-specific variables. $X'_{it}$ includes GDP growth rate, country-level productivity, total trade and the national spending on education, while $Z'_{sit}$ accounts for the sectoral productivity defined as the value added per employee and the sectoral average wage. These controls ensure that the impact of GVC participation on job creation by FDIs is accurately isolated by accounting for overall economic performance, efficiency, and human capital quality at the national level, as well as efficiency and labor cost variations within sectors.

$\alpha_s$ is the sector fixed effects to control for unobserved sector-specific factors like industry-specific regulations, technological advancements, production processes, and market structures; $\alpha_i$ is the country fixed effects to control for unobserved country-specific factors like national policies, legal systems, cultural factors, economic stability, and overall investment climate; and $\alpha_t$ is time fixed effects to control for unobserved time-specific factors like global economic conditions, technological innovations, and changes in international trade policies. The use of fixed effects addresses several issues, such as the potential concentration of GVC in industries with lower wages, and the impact of time-varying shocks on GVC and employment. Additionally, varying levels of openness among countries influence GVC involvement. $\varepsilon_{sit}$ is the error term capturing unobserved factors affecting FDI-induced job creation.

The model is well-suited for understanding whether GVC participation relates to FDI-induced job creation because it effectively separates the influence of GVC involvement from other factors that might affect employment. By controlling for broader economic conditions, sector-specific characteristics, and overall productivity, the model ensures a clear focus on how GVC intensity impacts job creation driven by foreign investment. This approach allows for a precise evaluation of whether countries with higher GVC engagement see more jobs created through FDI, aligning well with the study's objective.

### 3.2. Data and GVC measurement

We have on disposal micro data - individual FDI projects - i.e. FDI Markets Crossborder Monitor database, the most comprehensive online database for cross-border greenfield investments in the EU-27 – the geographic group we work with. The database originates from cross-border direct investment and contains all FDI projects from the years 2003 to 2023 that inflowed in the EU countries. For a given FDI project, it documents the country of origin and target country, the target sector, the target regions, the number of the number of jobs created, and the industry classifications. Therefore, in (1), we use measure the ln ($Jobs_{sit}$) through the number of jobs from the FDI Markets database; at the outset, we disregard the regional variation (since the other variables are either at the sectoral or national level, but we later allow for such variation as a form a robustness check. With respect to the sectoral identifier, the database disaggregates at a more detailed level than two-digit NACE Rev.2, but we use correspondence tables from Kordalska et al. (2022) and aggregated at the two-digit NACE Rev.2.

We measure GVC participation through their forward and backward components, as follows:

- $Forward\ GVC\ Participation_{sit} = \frac{V_{GVC}}{Va}$           (2)
- $Backward\ GVC\ Participation_{sit} = \frac{Y_{GVC}}{Y}$           (3)

Where: $V_{GVC}$ is the GVC-related components of value-added in exports; $Va$ is the total value added; $Y_{GVC}$ is the GVC-related component of final production; and $Y$ is the total final production; all at the sectoral level.

The relative position of a sector in a GVC is represented by the ratio of forward to backward GVC production length, whereby the average production length forward is calculated as the ratio of GVC-related domestic value-added to its induced gross output; and the average production length backward is calculated as the ratio of GVC-related foreign value-added to its induced gross output (Wang et al. 2017a). More specifically:

$$GVC\_pos_{sit} = \frac{\frac{V_{GVC}}{Y_{Domestic}}}{\frac{Y_{GVC}}{Y_{Foreign}}} \qquad (4)$$

When $GVC\_pos_{sit} > 1$, the sector is more forward-oriented in the GVC, indicating that it adds significant domestic value to exports relative to its reliance on foreign inputs. Conversely, the sector is more backward-oriented in the GVC, relying more heavily on foreign inputs relative to the domestic value it adds.

We make use of the Trade in Value Added (TiVA) dataset from the OECD. It provides a comprehensive framework for analyzing global trade by tracking the value added at each stage of production across different countries and industries. Unlike traditional trade statistics that measure gross trade flows, TiVA offers a more detailed view by breaking down the value of exports and imports into their constituent parts—showing how much value is added in each country before the final product reaches the consumer. From this dataset, we approximate the GVC-related components as follows: forward component by "Domestic value added in gross exports"; and backward component by "Gross imports of intermediate products" as shares in value added and gross production, respectively.

**Figure 1** reveals distinct patterns in GVC participation in EU-27 and its potential relation with jobs and output, from 2003 to 2020. The forward GVC participation forward (blue line) exhibits a steadiness before the 2008 financial crisis, which intensifies after the outbreak of the crisis, but stabilizes again afterwards. This movement in forward participation suggests a stable contribution to other countries' exports, which may have been contributing to both jobs and output growth (yellow line) during the pre-crisis period. However, the increase in forward GVC participation around the financial crisis correlates with a significant drop in both FDI-driven job creation and overall output. This counterintuitive outcome implies that increased domestic inputs in exports during and after periods of economic turmoil might not always lead to immediate or proportional benefits for employment and production, possibly due to the broader economic disruptions affecting demand, investment, and market stability during the crisis.

The backward GVC participation (orange line) reflects the degree to which the EU-27 relies on imported inputs for its production processes. Over the period from 2003 to 2020, the line indicates a relatively stable trend with slight fluctuations. Initially, backward GVC participation shows a minor increase, peaking slightly before the 2008 financial crisis, which suggests a growing reliance on imported inputs during the early 2000s. However, this trend does not exhibit the same dramatic shifts seen in forward GVC participation. After the financial crisis, the backward GVC participation stabilizes and then slightly decreases, indicating a potential reduction in the dependency on foreign inputs. This stability suggests that the EU-27's reliance on imported inputs has been relatively consistent over time. The slight decline post-crisis might indicate a shift towards more localized sourcing or a reduced need for imported inputs as the region adjusted to the economic challenges of the period. However, the relative stability also points to the persistent importance of global supply chains in the EU-27's production processes, even amidst broader economic disruptions.

**Figure 1 – Average GVC participation (forward and backward) and jobs generated by FDI**

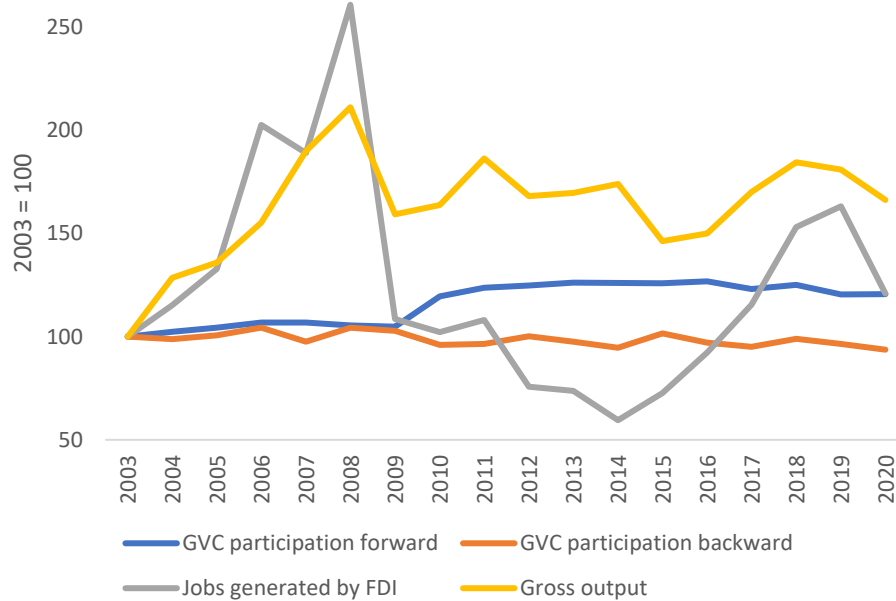

*Source: Authors' calculation.*

We further utilize the comprehensive database of GVC positioning measures of Mancini et al. (2024). Based on TiVA database (among others), they calculate measures of upstreamness of a country/sector - the distance of its productions from the final demand; and downstreamness - the distance from the factors of production (sources of value-added). Hence, $GVC\_pos_{sit}$ in (3) would be simply a ratio of the upstreamness and downstreamness measures. **Figure 2** presents the average GVC production chain length for the EU-27 in 2019. High values for both indices for counties like Luxembourg imply that they participate in long value chains, with many stages both backward and forward. By contrast, considered in both direction, lengths for countries like Slovenia or Greece are short. Some countries reveal differences in the two indices, for example, Cyprus is farther upstream, while Croatia is further downstream.

Figure 2 – Average GVC production length (forward and backward)

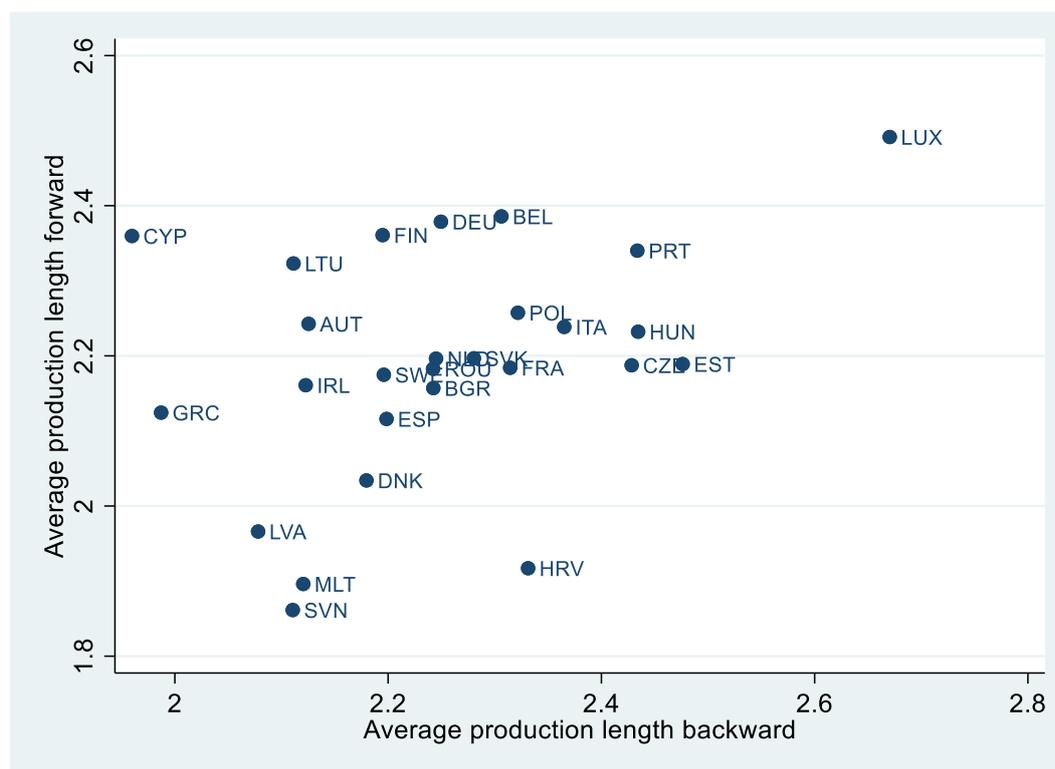

*Source:* GVC positioning database of Mancini et al. (2024).

The macro-level variables, namely: GDP growth, GDP per employed person and the government spending on education are derived from the World Bank's World Development Indicators. The sectoral variables, namely: the value added per workers and the average wage are derived from UNIDO's industrial database.

Given we integrate a variety of database, our final time period is 2003-2020 for the NACE Rev.2 sectors of agriculture, industry and construction. The FDI Market database is quite detailed and spans over 2003-2023. TiVA database contains virtually all two-digit NACE Rev.2 sectors (with a few aggregations, though) for the 2003-2020 period, while UNIDO and GVC positioning database provide info only for NACE Rev.2 codes of 1-43.

Full variable description and descriptive statistics is provided in the Annex (**Table A 1** and **Table A 2**). **Table A 3** presents inter-variables correlations, revealing the expectedly high correlation between sectoral productivity and sectoral wages; hence, we run them in separate regressions.

### 3.3. Why are FDI-generated jobs and GVC participation endogenous?

Crucial to estimating model (1) is to recognize the endogeneity between the GVC participation and jobs generated by incoming FDIs. First and foremost, there exists a bidirectional causality between FDI and job creation. When multinational companies invest in a country, they often establish new operations, expand existing ones, or develop new projects, all of which require labor, thus directly creating jobs. However, the opposite applies: the availability of a skilled labor force and job creation dynamics can influence the attractiveness of a country to foreign investors. Regions with higher employment rates and better workforce skills may draw more FDI due to the availability of human resources necessary for business operations (Kottaridi et al. 2019; Villaverde and Maza, 2015; Pantelopoulos, 2022; Miningou and Tapsoba, 2020).

A similar bidirectional relationship is observed between GVC participation and FDI. Countries and sectors deeply embedded in GVCs are often more attractive to foreign investors (Martínez-Galán and Fontoura, 2019; Carril-Caccia and Pavlova, 2019; Zhao, 2021). Participation in GVCs can signal efficiency, established networks, and a reliable supply chain, making a location more appealing for FDI (Amendolagine et al. 2019). On the other hand, FDI can enhance a country's GVC participation (Okah Efogo et al. 2022; Fernandes et al. 2022). Multinational companies often bring advanced technologies, managerial expertise, and access to international markets, which can integrate the host country more deeply into global production networks.

Omitted variable bias is another critical factor. Economic environment factors such as economic stability, institutional quality, infrastructure, and policies affect both FDI inflows and job creation (Sahiti et al. 2018). Government policies aimed at attracting FDI, such as tax incentives, subsidies, or regulatory reforms, can likewise simultaneously influence GVC participation (De Marchi and Alford, 2022). If these factors are not adequately controlled for, they can bias the estimation of the relationship between FDI-generated jobs and GVC participation. Our country fixed effects work to attenuate the endogeneity from this source. Additionally, different sectors have varying capacities for job creation and levels of GVC participation. High-tech industries may offer fewer but higher-paying jobs and have complex GVC networks, whereas labor-intensive industries might create more jobs with simpler value chains. Our industry fixed effects work to attenuate the endogeneity stemming from this source.

To accurately estimate the effects of GVC participation on FDI-induced job creation, it is essential to address these endogeneity concerns. This can be achieved by employing appropriate econometric techniques. In the empirical model outlined, the use of sector, country, and time fixed effects helps mitigate some of these concerns by controlling for unobserved heterogeneity. However, the inherent endogeneity between FDI-generated jobs and GVC participation necessitates the use of additional instrumental variables and robustness checks to ensure reliable and unbiased estimates. This comprehensive approach allows for a more accurate understanding of the dynamic interplay between FDI, GVC participation, and job creation.

### 3.4. Estimation method

To estimate the relationship between GVC participation and FDI-induced job creation, we employ the Generalized Method of Moments (GMM) estimator (Wooldridge, 2002). This method is particularly suitable for addressing potential endogeneity arising from bidirectional causality and omitted variable bias, which can lead to biased estimates if not properly accounted for. The GMM approach helps mitigate this by using instrumental variables (IVs), specifically lagged values of the endogenous variables and other exogenous instruments. These instruments are chosen to ensure they are uncorrelated with the error term, providing consistent estimates. This is particularly important given the panel nature of our data, which spans multiple countries, sectors, and years. By accounting for potential heteroskedasticity and autocorrelation, GMM provides more accurate standard errors and confidence intervals.

Key diagnostic tests are integral to our methodology. The weak identification test assesses whether the instruments are sufficiently correlated with the endogenous regressors, ensuring that they are strong instruments. Weak instruments can lead to biased estimates, making this test essential for validating our instrument choice. Additionally, Hansen's J test evaluates the validity of the instruments by checking their correlation with the error term, confirming their suitability for the model.

This approach ensures that our estimation method provides reliable insights into the dynamic interplay between GVC participation and FDI-induced job creation in the EU-27, offering a robust analysis of these complex economic interactions.

## 4. Results and discussion

### 4.1. Baseline results

The baseline results of model (1) are presented in **Table 1**. Columns (1)-(5) refer to the forward GVC participation and columns (6)-(10) to the backward one, adding explanatory variables group by group. Towards the bottom of the table, underidentification and Hansen J tests confirm the appropriateness of the instruments used and the validity of the models.

Results suggest that forward GVC forward participation positively impacts FDI-generated jobs. An increase in domestic value-added in exports – referring to more intense forward GVC participation, leads to a significant rise in FDI-related job creation, showing that deeper integration into the export side of the GVC causes more jobs via FDIs. By adding more value domestically before exporting, sectors likely become more attractive to foreign investors who seek efficiency and established supply networks. The positive employment effects arise from enhanced production capacity, local supplier linkages, and the overall economic dynamism that appeals to FDIs.

Notably, though, this impact reduces when sectoral productivity and sector-specific characteristics are included; and almost halves when sectoral wage levels are included instead. These factors might capture some of the effects that were initially attributed to GVC participation alone. For example, higher productivity or wages in certain sectors might independently attract FDI – as evidenced by the positive and significant coefficient on these sectoral variables, reducing the marginal impact of GVC participation. This suggests that forward GVC participation's effect on jobs is partly mediated by these sector-specific characteristics, which absorb part of the initial impact.

In contrast, backward GVC participation (columns 6-10) shows a negative effect on job creation. Sectors and countries relying more on foreign inputs experience lower volume of FDI-generated jobs, implying that outsourcing or lower domestic labor requirements reduce FDI's job-creating potential; this reduces the need for domestic labor as more production stages are completed abroad. Additionally, relying on foreign inputs may signal less competitive domestic industries, making them less attractive for FDIs that seek robust local supply chains. Thus, instead of creating jobs, such sectors may see a lower employment due to lower domestic production requirements and reduced attractiveness for job-generating investments.

The role of backward and forward linkages within the labor market is consistent with Taglioni and Winkler (2016), who show that GVC participants experience different effects depending on whether they act as buyers or sellers. Our findings are consistent with the analyses of Szymczak and Wolszczak-Derlacz (2021) and Mingyang et al. (2023), who also confirm that developed countries increase employment by improving forward participation in GVCs. We confirm the finding of Farole et al. (2018) that forward participation in GVCs tends to be biased towards skilled labor, such that in developed countries with a large abundance of skilled labor, the relationship between job creation and forward participation tends to be positive. Moreover, our results support the negative relationship between participation in backward GVCs and employment that is often observed in developed countries (Farole 2016). However, we cannot confirm the positive "expansion effect" (stronger backward participation leads to an expansion of industry and thus to an increase in labor demand) observed by Pan (2020) for the American economy and by Szymczak and Wolszczak-Derlacz (2021) for 46 middle- and high-income countries. In the case of the EU economies, we observed a strong substitution effect.

However, it should not be neglected that the effect of backward GVC participation on jobs is lost when sectoral wages and invariant sector characteristics are included in **Table 1**; more notably, in this specification trade gains significance, revealing the usual results that higher trade leads to higher employment. This shift may indicate that once sectoral wages are accounted for, the job-reducing effects of backward GVC participation are overshadowed by the broader positive impact of increased trade on employment. The inclusion of wages may also reveal complex interactions with trade. This could imply that while reliance on foreign inputs might reduce domestic labor demand in isolation, the overall economic benefits of higher trade activity—such as increased demand for goods and services—outweigh these effects when sectoral wages are favorable. Such development may be expected by the 'new' new trade theory (Feenstra and Hanson, 1996; Marjit et al. 2024).

The negative GVC position coefficients indicate that FDI in upstream sectors—those further from final demand—creates fewer jobs compared to investment in downstream sectors. Sectors closer to final demand, which involve final assembly or consumer services, have higher labor needs and therefore are more effective at generating employment. This suggests that for maximizing job creation through FDI, targeting sectors nearer to the end consumer is more advantageous. For example, upstream sectors might be more capital-intensive or conduct highly specialized processes or have higher sectoral productivity, which have been accounted for in the initial model.

As already evidenced, as additional explanatory variables are included, the coefficients for GVC participation change. These variables also offer further insights: sectors with higher productivity and higher wages attract more FDI-generated jobs, and lower country productivity negatively impacts FDI job creation. Higher wages can attract more FDI-generated jobs by signaling a skilled labor force, greater market potential, and economic stability. However, higher country productivity might negatively impact job creation because it often leads to more capital-intensive and efficient production processes, reducing the number of new jobs created. In essence, while higher wages can attract FDI by offering a more attractive labor market, higher overall productivity can lead to fewer jobs as businesses invest in technology and efficiency improvements instead of expanding their workforce.

Table 1 – Baseline results

| | Dependent variable: Log of number of jobs created by FDIs | | | | | | | | | |
|---|---|---|---|---|---|---|---|---|---|---|
| | Forward participation | | | | | Backward participation | | | | |
| | (1) | (2) | (3) | (4) | (5) | (6) | (7) | (8) | (9) | (10) |
| GVC participation | 0.941*** | 0.959*** | 0.994*** | 0.845*** | 0.536*** | -1.164*** | -1.145*** | -1.130*** | -0.866*** | -0.156 |
| | (0.088) | (0.088) | (0.090) | (0.104) | (0.116) | (0.140) | (0.139) | (0.134) | (0.177) | (0.212) |
| GVC position | | -0.816*** | -0.729*** | -0.740*** | -0.671*** | | -0.528*** | -0.453** | -0.466** | -0.449* |
| | | (0.194) | (0.195) | (0.229) | (0.227) | | (0.197) | (0.197) | (0.230) | (0.226) |
| GDP growth rate | | | 0.0264 | 0.00797 | 0.00817 | | | 0.035 | 0.00719 | -0.015 |
| | | | (0.042) | (0.042) | (0.041) | | | (0.042) | (0.042) | (0.041) |
| Log of GDP per person employed | | | -1.716* | -2.862*** | -2.821*** | | | -1.263 | -2.699*** | -3.213*** |
| | | | (0.883) | (0.828) | (0.818) | | | (0.890) | (0.813) | (0.815) |
| Government spending on education | | | 0.0996 | 0.207* | 0.184 | | | 0.104 | 0.167 | 0.142 |
| | | | (0.103) | (0.118) | (0.116) | | | (0.101) | (0.116) | (0.115) |
| Trade | | | 0.009 | 0.00912 | 0.00799 | | | 0.007 | 0.0112* | 0.00988* |
| | | | (0.007) | (0.007) | (0.006) | | | (0.007) | (0.006) | (0.006) |
| Log of sectoral productivity | | | | 0.206* | | | | | 0.183 | |
| | | | | (0.121) | | | | | (0.127) | |
| Log of sectoral average wage | | | | | 0.297*** | | | | | 0.415*** |
| | | | | | (0.054) | | | | | (0.061) |
| Constant | 2.583*** | 3.343*** | 22.12** | 32.35*** | 28.17*** | 3.114*** | 3.609*** | 17.28* | 30.36*** | 30.02*** |
| | (0.201) | (0.278) | (9.954) | (9.552) | (9.327) | (0.198) | (0.276) | (10.050) | (9.359) | (9.185) |
| Observations | 3424 | 3424 | 3339 | 2586 | 2598 | 3451 | 3451 | 3366 | 2629 | 2641 |
| R-squared | 0.507 | 0.509 | 0.517 | 0.522 | 0.534 | 0.506 | 0.507 | 0.514 | 0.521 | 0.534 |
| Underidentification test (Kleibergen-Paap rk LM statistic, p-value): | 0.0000 | 0.0000 | 0.0000 | 0.0000 | 0.0000 | 0.0000 | 0.0000 | 0.0000 | 0.0000 | 0.0000 |
| Hansen J statistic (overidentification test of all instruments, p-value): | 0.397 | 0.476 | 0.679 | 0.455 | 0.586 | 0.39 | 0.361 | 0.677 | 0.533 | 0.622 |

Source: Author's estimations.
Note: *, ** and *** refer to statistical significance at the 10, 5 and 1% level, respectively. Robust standard errors provided in parentheses. Regressions include sector, country and time dummies, all not reported but available on request.

### 4.2. Results with some regional variation

Our FDI Markets Crossborder Monitor database contains one more favorable information, that is, the within-country region in which the FDI inflowed. While we have no on disposal other variables at the regional, we could still make some use of this regional variation, i.e. in (1) allowing ln ($Jobs_{rsit}$) adding a subscript $r$ to denote the region. This still requires some aggregation in our FDI data because there could be multiple FDIs inflowing in the same sector, country and year. Results serve robustness check and are presented in **Table 2**. The pattern of results is similarly as in the baseline results: intensity of the forward GVC participation leads to more FDI-generated employment in the recipient country-industry-region triple; while the intensity of the backward GVC participation leads to fewer jobs, if at all significant. It is to be noted, however, that the coefficients on the forward GVC participation are now slightly smaller (in a largely significant fashion), which is not the case for the coefficients on the backward GVC participation. This suggests that once regional variation is factored in, the role of the domestic component in exports for job creation diminishes. This important result may have a couple of lines of explanation.

First, some regions may be more capable of leveraging foreign inputs for job creation due to better infrastructure, skilled labor, or more competitive industries. When these regional differences are factored in, the overall effect of forward GVC participation on job creation may appear smaller because the positive impact is not uniformly distributed across regions. Second, the domestic component in exports, which forward GVC participation represents, may become less critical for job creation when the analysis zooms in on regional differences. This could be because regional factors, such as local policies, labor market conditions, or specific industries, may influence job creation more than the general level of domestic input in exports.

On the other hand, while GVC position further indicates that FDIs in upstream sectors create fewer jobs compared to investment in downstream sectors, the coefficient slightly intensifies in the case of the backward linkages model (though, a difference at the verge of significance). The slight intensification of the effect in the backward linkages model suggests that when a sector in particular region relies more on foreign inputs (as in backward GVC participation), the already lower job creation potential of FDI in upstream sectors may be further reduced possibly capturing regional varieties related to capital intensity or policies related to attraction of high-profile FDIs.

The same pattern of coefficients and their movement as in the baseline specification (**Table 1**) is obtained when, instead, number of affected regions by FDI inflows is directly controlled for in the regression (

Table 3). The coefficient on the number of affected regions suggests that the more regions affected, the stronger the job creation by FDI, no matter the other controls involved in the regression. This implies that the more regions benefiting from FDI, the greater the overall employment impact, regardless of other factors like sectoral productivity or wages. The robustness of this relationship suggests that spreading FDI across multiple regions mitigates some of the constraints that may limit job creation in more concentrated areas or specific sectors. Moreover, this regional diversification of FDI inflows could contribute to more inclusive economic growth by stimulating job creation in areas previously less integrated into global value chains. The results highlight that a wider distribution of FDI helps strengthen local economies, fostering greater employment opportunities across the country and creating positive linkages between regional and global markets.

Table 2 – Results – regional variation

| | Dependent variable: Log of number of jobs created by FDIs | | | | | | | | | |
|---|---|---|---|---|---|---|---|---|---|---|
| | Forward participation | | | | | Backward participation | | | | |
| | (1) | (2) | (3) | (4) | (5) | (6) | (7) | (8) | (9) | (10) |
| GVC participation | 0.641*** | 0.663*** | 0.697*** | 0.623*** | 0.415*** | -1.026*** | -1.010*** | -1.038*** | -0.751*** | -0.183 |
| | (0.091) | (0.091) | (0.092) | (0.108) | (0.118) | (0.126) | (0.125) | (0.126) | (0.147) | (0.180) |
| GVC position | | -0.751*** | -0.716*** | -0.421* | -0.460* | | -0.581*** | -0.546*** | -0.321 | -0.399* |
| | | (0.203) | (0.205) | (0.239) | (0.238) | | (0.203) | (0.205) | (0.238) | (0.236) |
| GDP growth rate | | | 0.0124 | -0.013 | -0.00666 | | | 0.00887 | -0.0123 | -0.0212 |
| | | | (0.042) | (0.041) | (0.039) | | | (0.039) | (0.039) | (0.038) |
| Log of GDP per person employed | | | -0.531 | -2.117** | -2.049** | | | -0.498 | -1.970** | -2.257*** |
| | | | (0.925) | (0.889) | (0.872) | | | (0.853) | (0.828) | (0.825) |
| Government spending on education | | | 0.0572 | 0.163 | 0.179 | | | 0.0495 | 0.138 | 0.17 |
| | | | (0.107) | (0.123) | (0.122) | | | (0.103) | (0.120) | (0.119) |
| Trade | | | 0.0058 | 0.0109 | 0.00952 | | | 0.00492 | 0.0105* | 0.0095 |
| | | | (0.008) | (0.007) | (0.007) | | | (0.007) | (0.006) | (0.006) |
| Log of sectoral productivity | | | | 0.14 | | | | | 0.165 | |
| | | | | (0.121) | | | | | (0.129) | |
| Log of sectoral average wage | | | | | 0.211*** | | | | | 0.304*** |
| | | | | | (0.052) | | | | | (0.062) |
| Constant | 2.643*** | 3.357*** | 8.68 | 23.55** | 20.11** | 2.935*** | 3.493*** | 8.561 | 21.62** | 20.57** |
| | (0.277) | (0.343) | (10.450) | (10.360) | (10.060) | (0.266) | (0.334) | (9.725) | (9.642) | (9.446) |
| Observations | 4,737 | 4,737 | 4,606 | 3,636 | 3,669 | 4,788 | 4,788 | 4,656 | 3,675 | 3,707 |
| R-squared | 0.386 | 0.387 | 0.395 | 0.35 | 0.357 | 0.409 | 0.409 | 0.417 | 0.363 | 0.37 |
| Underidentification test (Kleibergen-Paap rk LM statistic, p-value): | 0.000 | 0.000 | 0.000 | 0.000 | 0.000 | 0.000 | 0.000 | 0.000 | 0.000 | 0.000 |
| Hansen J statistic (overidentification test of all instruments, p-value): | 0.411 | 0.569 | 0.738 | 0.603 | 0.541 | 0.444 | 0.365 | 0.707 | 0.635 | 0.411 |

Source: Author's estimations.
Note: *, ** and *** refer to statistical significance at the 10, 5 and 1% level, respectively. Robust standard errors provided in parentheses. Regressions include region, sector, country and time dummies, all not reported but available on request.

Table 3 – Results – regional variation (2)

| | Dependent variable: Log of number of jobs created by FDIs | | | | | | | | | |
|---|---|---|---|---|---|---|---|---|---|---|
| | Forward participation | | | | | Backward participation | | | | |
| | (1) | (2) | (3) | (4) | (5) | (6) | (7) | (8) | (9) | (10) |
| GVC participation | 0.900*** | 0.918*** | 0.968*** | 0.819*** | 0.556*** | -1.047*** | -1.034*** | -1.054*** | -0.766*** | -0.162 |
| | (0.111) | (0.110) | (0.115) | (0.137) | (0.153) | (0.146) | (0.144) | (0.138) | (0.197) | (0.235) |
| GVC position | | -0.793*** | -0.689*** | -0.599** | -0.606** | | -0.575** | -0.489** | -0.361 | -0.403 |
| | | (0.228) | (0.227) | (0.254) | (0.253) | | (0.228) | (0.226) | (0.253) | (0.250) |
| GDP growth rate | | | 0.0647 | 0.0359 | 0.0295 | | | 0.062 | 0.0217 | -0.00402 |
| | | | (0.048) | (0.045) | (0.043) | | | (0.048) | (0.044) | (0.043) |
| Log of GDP per person employed | | | -0.543 | -2.101** | -2.083** | | | -0.324 | -2.165** | -2.653*** |
| | | | (1.119) | (1.016) | (0.989) | | | (1.119) | (0.992) | (0.976) |
| Government spending on education | | | 0.0791 | 0.199 | 0.173 | | | 0.0599 | 0.132 | 0.111 |
| | | | (0.111) | (0.132) | (0.129) | | | (0.109) | (0.129) | (0.128) |
| Trade | | | 0.0018 | 0.00612 | 0.00578 | | | 0.000764 | 0.00905 | 0.00914 |
| | | | (0.009) | (0.008) | (0.007) | | | (0.009) | (0.007) | (0.007) |
| Log of sectoral productivity | | | | 0.129 | | | | | 0.129 | |
| | | | | (0.133) | | | | | (0.147) | |
| Log of sectoral average wage | | | | | 0.225*** | | | | | 0.368*** |
| | | | | | (0.068) | | | | | (0.075) |
| Number of regions affected | 0.170*** | 0.163*** | 0.156*** | 0.122*** | 0.102*** | 0.144*** | 0.141*** | 0.135*** | 0.111*** | 0.0853*** |
| | (0.024) | (0.024) | (0.025) | (0.027) | (0.028) | (0.025) | (0.025) | (0.025) | (0.028) | (0.028) |
| Constant | 3.027*** | 3.801*** | 9.792 | 24.78** | 21.51* | 3.481*** | 4.049*** | 7.674 | 25.26** | 24.87** |
| | (0.338) | (0.419) | (12.650) | (11.770) | (11.480) | (0.334) | (0.412) | (12.690) | (11.470) | (11.180) |
| Observations | 1,983 | 1,983 | 1,965 | 1,521 | 1,529 | 2,014 | 2,014 | 1,996 | 1,550 | 1,558 |
| R-squared | 0.56 | 0.562 | 0.561 | 0.566 | 0.575 | 0.56 | 0.561 | 0.56 | 0.571 | 0.579 |
| Underidentification test (Kleibergen-Paap rk LM statistic, p-value): | 0.000 | 0.000 | 0.000 | 0.000 | 0.000 | 0.000 | 0.000 | 0.000 | 0.000 | 0.000 |
| Hansen J statistic (overidentification test of all instruments, p-value): | 0.42 | 0.38 | 0.617 | 0.511 | 0.565 | 0.534 | 0.458 | 0.714 | 0.652 | 0.452 |

Source: Author's estimations.
Note: *, ** and *** refer to statistical significance at the 10, 5 and 1% level, respectively. Robust standard errors provided in parentheses. Regressions include sector, country and time dummies, all not reported but available on request.

### 4.3. Sectoral and geographic heterogeneity

We examine geographic and sectoral heterogeneity in our main results: GVC participation and position. For the geographic division, we use the usual disaggregation of old and new EU member states, the dividing line being the countries who joined before and after 2004. With respect to the sectoral heterogeneity, we group the two-digit NACE Rev.2 codes around two concepts: labor/capital intensity, following Kucera and Sarna (2006) though with some adjustment; and technological intensity, based on OECD (2011) and Eurostat (2024).

Results are presented in **Table 4**; due to space, only the main coefficients of interest are presented. The upper part of the table provides insights into how forward GVC participation and GVC position influence employment outcomes across different geographic and sectoral divisions. Starting with the geographic division between old and new EU member states, the results indicate a clear difference in how forward GVC participation impacts job creation. In the old EU member states, forward GVC participation is associated with significantly higher employment, suggesting that these countries are better positioned to benefit from their integration into global value chains. This could be due to their more established industries, stronger infrastructure, and deeper integration into international markets, which enable them to convert GVC participation into job growth (Pan, 2020). In contrast, in the new EU member states, the relationship between forward GVC participation and employment is positive but not statistically significant. This implies that these newer member states may still be developing the capabilities needed to fully capitalize on GVC participation for job creation, reflecting perhaps less mature industries or weaker linkages with global markets. We recognize the latter is more heterogenous group, which may be also a driver of the result. Our findings are in line with those of Stollinger (2019), who notes a specialization of the old EU members primarily in high value-added services (such as R&D, marketing management), which account for a growing share of global trade. This leads to strong forward linkages in the old EU-15 members and can influence employment growth in the service sectors.

Looking at the sectoral divisions, the results show that forward GVC participation has different effects depending on the nature of the industry. In capital-intensive sectors, forward GVC participation strongly boosts employment, likely because these sectors benefit from integrating into global markets where they can leverage their capital assets. This is in contrast to labor-intensive sectors, where the effect of forward GVC participation on employment is not significant. This may suggest that labor-intensive sectors are less able to translate participation in global value chains into job growth, possibly due to competition from lower-wage countries or challenges in scaling labor in response to global demand.

When breaking down the data by technology level, forward GVC participation is most effective in creating jobs in medium-tech sectors, followed closely by high-tech sectors. This highlights the importance of technological capabilities in benefiting from global value chains. Medium- and high-tech sectors likely have the innovation, skilled labor, and production capabilities necessary to take full advantage of GVC participation. Our results confirm the findings of Bontadini et al. (2021), who find that countries that are leaders in terms of technological intensity have greater integration in GVCs and a greater share of employment in headquarters functions (strategic planning, corporate communications, tax, legal, marketing, finance, human resources). In contrast, low-tech sectors do not see a significant employment boost from forward GVC participation, possibly because these industries face stiff competition from other countries and may not have the same capacity to scale up employment in response to global demand.

When it comes to GVC position, which reflects whether a country is upstream or downstream in the value chain, the result is insignificant for the old EU member states, when backward GVC participation is controlled for. This could suggest that these countries have the capacity to maintain employment levels regardless of their position in the value chain, possibly due to diversified economies. Our findings are in line with the analysis by Stollinger (2019), who identifies a specialization of the old EU countries primarily in services with high added value, which belong to upstream activities (such as R&D) and downstream activities (such as marketing management). However, in the new EU member states, the negative and significant coefficient suggests that being upstream in the value chain is associated with fewer jobs (as in the overall results in **Table 1**). Stollinger (2019) and Kordalska and Olczyk (2022) find that new EU member states specialize in middle-production-process tasks, far from final demand. In capital-intensive and medium-tech sectors, being upstream in the value chain significantly reduces employment. This could be because these sectors, when positioned earlier in the production process, are more reliant on automation and less on labor, leading to fewer job opportunities. In labor-intensive and low-tech sectors, the relationship is not significant, perhaps because these sectors are already operating at a low technological level where the position in the value chain does not drastically affect job creation (Antras, 2020).

The coefficients presented for backward GVC participation – in the lower part of **Table 4** - provide a different perspective on how engagement in global value chains affects employment outcomes across various geographic and sectoral divisions. Backward GVC participation in the old EU member states does not matter for employment. This could imply that these countries have sufficient domestic capacity and competitive industries that mitigate the potential negative effects of relying on foreign inputs. In the new EU member states, however, the coefficient for backward GVC participation is negative and significant, as in the baseline results. New EU member states, which are still building their industrial capabilities, may find that importing inputs limits their ability to create jobs domestically, as more of the production process is handled abroad. This could reflect weaker domestic industries that are less able to compete or substitute foreign inputs with local production. The results of our study confirm the functional dichotomy between new and old EU countries observed by Kordalska and Olczyk (2022). The authors emphasize the specialization of the new countries in tasks related to production and especially assembly, which implies strong backward linkages on the one hand and the increasing automation of routine activities on the other could have a negative impact on national employment.

Looking at the sectoral divisions, the results for backward GVC participation provide a complex picture. In labor-intensive sectors, the negative and non-significant coefficient suggests that reliance on foreign inputs does not notably affect employment, which may indicate that these sectors are less dependent on backward GVC participation to create jobs. However, in capital-intensive sectors, the impact is similarly non-significant, but with a very small negative coefficient. This might suggest that capital-intensive sectors do not benefit from backward GVC participation and may even experience slight job losses, as more production is outsourced abroad.

The results in technological division reveal that backward GVC participation has a significant negative impact on employment in low-tech sectors only. This suggests that in industries with low technological content, relying on foreign inputs substantially reduces job creation, possibly because these sectors struggle to compete with cheaper or more efficient foreign producers. In high- and medium-tech sectors, while the coefficient is negative, it is not significant, indicating that these sectors may be better able to integrate foreign inputs without losing jobs.

The impact of GVC position, when backward GVC participation is controlled for, shows a similar result for the new MS. For the rest of disaggregations, the coefficient is statistically insignificant, though consistently negative, suggesting that sectors positioned upstream in the value chain may still experience some reduction in employment, even if the effect is not strong enough to be statistically confirmed across different sectoral or geographic divisions. This could indicate that the challenges associated with being upstream, such as reliance on capital-intensive processes or reduced demand for domestic labor, persist across various contexts, albeit to a lesser extent.

Table 4 – Results – sectoral and geographic variation

|  | Geographic division | | Sectoral division (1) | | Sectoral division (2) | | |
| --- | --- | --- | --- | --- | --- | --- | --- |
|  | Old EU-MS | New EU-MS | Labor intensive | Capital intensive | High tech | Medium tech | Low tech |
|  | Forward GVC participation | | | | | | |
| GVC participation | 0.437*** | 0.154 | 0.179 | 0.643*** | 0.578*** | 0.797*** | -0.136 |
|  | (0.155) | (0.233) | (0.349) | (0.134) | (0.178) | (0.252) | (0.389) |
| GVC position | -0.165 | -0.610** | -0.533 | -0.553** | -0.079 | -1.329*** | -0.443 |
|  | (0.397) | (0.272) | (0.883) | (0.247) | (0.309) | (0.514) | (1.030) |
|  | Backward GVC participation | | | | | | |
| GVC participation | 0.224 | -0.652** | -1.56 | -0.093 | -0.0046 | -0.645 | -2.118* |
|  | (0.288) | (0.278) | (1.020) | (0.236) | (0.312) | (0.442) | (1.133) |
| GVC position | -0.0191 | -0.471* | -0.979 | -0.29 | -0.0217 | -0.481 | -1.265 |
|  | (0.389) | (0.268) | (0.877) | (0.246) | (0.313) | (0.481) | (1.057) |

Source: Author's estimations.
Note: *, ** and *** refer to statistical significance at the 10, 5 and 1% level, respectively. Robust standard errors provided in parentheses. Regressions include controls, sector, country and time dummies, all not reported but available on request.

## 5. Conclusions and policy implications

This study examined the varying effects of global value chain (GVC) participation on foreign direct investment (FDI)-related job creation within the EU-27, employing a labor demand function. The model accounts for sector-specific and regional factors by incorporating sector, country, and time fixed effects to address potential biases and endogeneity issues. By leveraging data from the FDI Markets Crossborder Monitor and employing the Generalized Method of Moments (GMM) estimator, the study provides a robust analysis of the dynamic relationship between GVC participation and job creation. It covers EU-27 members states, over the 2003-2020 period, and agriculture, industry and construction sectors.

Results indicate that forward GVC participation significantly boosts FDI-related job creation, as deeper integration into global value chains (GVCs) on the export side attracts foreign investors by enhancing domestic value-added and production capacity. However, this positive effect is diminished when sector-specific factors like productivity and wages are included, suggesting that these characteristics mediate the impact of GVC participation. In contrast, backward GVC participation, characterized by reliance on foreign inputs, generally reduces FDI-generated jobs due to lower domestic labor requirements and diminished competitiveness. The negative impact of being upstream in the value chain further highlights the challenges of job creation in sectors distant from final demand, where capital intensity and efficiency often reduce employment. Nevertheless, ultimately, the negative effect of backward GVC participation on jobs becomes insignificant, while trade gains importance, underscoring the broader positive impact of trade on employment when sectoral dynamics are considered.

The analysis incorporating regional variation in FDI inflows – *inter alia* to serve for robustness check, reveals that forward GVC participation continues to positively impact FDI-generated employment, although its effect diminishes slightly when regional factors are considered. This suggests that regional characteristics, such as infrastructure and skilled labor, potentially play a significant role in job creation. In contrast, backward GVC participation continues to negatively impact employment, particularly in regions that rely more on foreign inputs, and this effect intensifies when considering upstream sectors. However, the negative impact of backward GVC participation on job creation is mitigated when FDI inflows are spread across multiple regions, indicating that regional diversification can help absorb and counterbalance some of the adverse effects.

Results suggest that forward GVC participation and GVC position have varying impacts on employment depending on the country's EU membership status and the nature of the sector. Old EU member states and capital- or medium-to-high-tech-intensive sectors seem to benefit most from forward GVC participation, while new EU member states and upstream positions in the value chain face more challenges in leveraging these global linkages for job growth. On the other hand, backward GVC participation appears to have a more adverse effect on employment, particularly in new EU member states and low-tech sectors. The reliance on foreign inputs seems to undermine job creation, especially in contexts where domestic industries are not competitive enough to benefit from participation in global value chains. This is further compounded by the negative impact of being upstream in the value chain, which seems to reduce employment, particularly in less advanced economies and sectors that are more capital-intensive.

The analysis underscores the nuanced impact of global value chain (GVC) participation on job creation, with significant policy implications. Forward GVC participation, which enhances domestic value-added and production capacity, positively influences FDI-related job creation. However, its effectiveness is moderated by sector-specific factors such as productivity and wages, and slightly diminished by regional characteristics. Therefore, policies should aim to bolster productivity and wage levels in sectors poised for GVC integration to fully leverage its benefits. Conversely, backward GVC participation tends to reduce job creation due to reliance on foreign inputs and capital-intensive processes, particularly in less advanced economies and low-tech sectors. This suggests that policies should focus on improving domestic competitiveness in these EU countries having the upstream position in GVCs.

The results show that EU countries cannot pursue the same policy of job creation through greater integration into GVC. Our study clearly shows that only greater forward participation leads to employment growth, and this relationship is only significant in the case of the old EU countries. The EU15 countries should aim to strengthen their current position in GVCs. This is not only about becoming more competitive in higher value-added activities, but also about involving more local actors, both companies and workers. Particular attention should be paid to the business services sector, which, according to Winkler et al. (2023), has been characterized by a growing number of jobs in all countries in recent years. Strengthening the position in higher value-added activities is a challenge for the old EU countries in the context of the simultaneous implementation of the Green Deal strategy and other Sustainable Development Goals.

The new EU countries face a much greater challenge, because the increasing forward and backward linkages do not lead to an increase in jobs. The new EU countries need to focus primarily on reducing backward participation by building a network of domestic suppliers. Policy makers should focus on developing linkages between GVCs and the domestic economy, often associated with greater diffusion of knowledge, technology and know-how from foreign investors or trading partners, together with strengthening the absorptive capacity of domestic firms (Taglioni and Winkler, 2016). Second, CEE countries should focus on upgrading GVCs by changing their functional specialization patterns and adopting new downstream activities in GVCs such as marketing or management. Specializing in these high value-added tasks involves high technologies and high levels of R&D, which has a positive spillover on technological upgrading (Elshaarawy and Ezza, 2023). It is not an easy task, because the CEE have determined region's role in the international division of labor. They focus on fabrication (assembly) activities, particularly in the automotive and electronics sectors with deep-rooted supplier networks that cannot be easily relocated.

Our analysis is a starting point for further research. It seems interesting to investigate whether the job creation driven by FDI and fostered by forward participation is accompanied by economic growth in the EU. According to Benetrix et al. (2023), GVCs development has completely changed the nature of FDIs and their potential impact on economic growth. On one hand, GVCs could limit "a country's capacity" to receive FDI; on the other, GVCs allow MNEs to employ low-wage workers in poor countries while maintaining high value-added production processes in countries with higher skill levels.

# References

Adarov, A., 2021. Interactions between global value chains and foreign direct investment: A network approach. WIIW Work. Pap. 204, 1–32.

Aleman-Castilla, B., 2020. Trade and labor market outcomes: Theory and evidence at the firm and worker levels. ILO Work. Pap. 12, 1–53.

Amendolagine, V., Presbitero, A.F., Rabellotti, R., Sanfilippo, M., 2019. Local sourcing in developing countries: The role of foreign direct investments and global value chains. World Dev. 113, 73–88. https://doi.org/10.1016/j.worlddev.2018.08.010

Antràs, P., 2020. Conceptual Aspects of Global Value Chains. World Bank Econ. Rev. 34, 551–574. https://doi.org/10.1093/wber/lhaa006

Bai, S., Zhang, B., Ning, Y., 2024. Measuring employment in global value chains based on an inter-country input-output model with multinational enterprises. Struct. Chang. Econ. Dyn. 68, 148–162.

Baldwin, R., 2011. Trade And Industrialisation After Globalisation's 2nd Unbundling: How Building And Joining A Supply Chain Are Different And Why It Matters. NBER Work. Pap. No. 17716, 1–39.

Bamber, P., Staritz, C., 2016. The gender dimensions of global value chains. Geneva: Int. Cent. Trade Sustain. Dev.

Banga, K., 2017. Impact of linking into global value chains on Indian employment. CEP Work. Pap. 2017/1, 1–33.

Benetrix, A., Pallan, H., Panizza, U., 2022. The Elusive Link Between FDI and Economic Growth. Policy Res. Work. Pap. 10422, 1–52.

Bernard, A., Jensen, J., Redding, S., Schott, P., 2007. Firms in International Trade. J. Econ. Perspect. 21, 105–130. https://doi.org/10.1257/jep.21.3.105

Bontadini, F., Evangelista, R., Meliciani, V., Savona, M., 2021. Asymmetries in Global Value Chain Integration, Technology and Employment Structures in Europe: Country and Sectoral Evidence, European Trade Union Institute. https://doi.org/10.2139/ssrn.3975956

Cadestin, C., De Backer, K., Desnoyers-James, I. Miroudot, S., Ye, M., Rigoi, D., 2018. Multinational enterprises and global value chains: New Insights on the trade-investment nexus. OECD Sci. Technol. Ind. Work. Pap. 2018/05, 1–36.

Camarero, M., López-Villavicencio, A. and Tamarit, C., 2022. Global value chains and unemployment in the EU – New insights on the role of the workforce composition and financial frictions, Publications Office of the European Union.

Cao, Q., Mohiuddin, M., Li, B., 2022. Impact of global value chain participation on rural migrant labourers' employment opportunities in urban areas in developing countries. Eur. J. Int. Manag. 1, 1–21. https://doi.org/10.1504/ejim.2022.10049626

Caraballo, J.G., Jiang, X., 2016. Value-Added Erosion in Global Value Chains: An Empirical Abebment. J. Econ. Issues 50, 288–296. https://doi.org/10.1080/00213624.2016.1148991

Carneiro, S., Neves, P.C., Afonso, O., Sochirca, E., 2024. Meta-analysis: global value chains and employment. Appl. Econ. 56, 2295–2314. https://doi.org/10.1080/00036846.2023.2186365

Carpa, N., Martínez-Zarzoso, I., 2022. The impact of global value chain participation on income inequality. Int. Econ. 169, 269–290. https://doi.org/10.1016/j.inteco.2022.02.002

Carril-Caccia, F. Pavlova, E., 2019. Mergers and acquisitions & trade: A global value chain analysis. World Econ. 43, 586–614. https://doi.org/10.1111/twec.12882

Chepeliev, M., Maliszewska, M., Osorio-Rodarte, I., Pereira, M.F.S.E., Mensbrugghe, D.V.D., 2022. Pandemic, climate mitigation, and reshoring: impacts of a changing global economy on trade, incomes, and poverty, World Bank Working Paper.

De Marchi, V., Alford, M., 2022. State policies and upgrading in global value chains: A systematic literature review. J. Int. Bus. Policy 5, 88–111. https://doi.org/10.1057/s42214-021-00107-8

Dine, M.N., 2019. Impact of global value chains' participation on employment in Turkey and spillovers effects. J. Econ. Integr. 34, 308–326. https://doi.org/10.11130/jei.2019.34.2.308

Duarte, A.E.S., Castilho, M. dos R., 2024. Does the Nature of Integration into GVC Matter for Job Creation? Evidence From Different Groups of Countries and Industries, Preprint.

Efogo, O.F., Wonyra, K.O., Osabuohien, E., 2022. Foreign direct investment and participation of developing countries in global value chains: lessons from the last decade. Int. Rev. Appl. Econ. 36, 264–284. https://doi.org/10.1080/02692171.2021.1962255

Elshaarawy, R., Ezzat, R.A., 2023. Global value chains, financial constraints, and innovation, Small Business Economics. Springer US. https://doi.org/10.1007/s11187-022-00685-8

Escaith, H., Inomata, S., Miroudot, S., 2018. Evolution of production networks in the Asia–Pacific region: A vision in value-added and employment dimensions, in: Asian Economic Integration in an Era of Global Uncertainty. pp. 155–183. https://doi.org/10.22459/aeiegu.01.2018.06

Eurostat, 2024. High-tech industry and knowledge-intensive services [WWW Document]. https://ec.europa.eu/eurostat/cache/metadata/en/htec_esms.htm.

Farole, T., 2016. Do global value chains create jobs? IZA World Labor, 1–11. https://doi.org/10.15185/izawol.291

Farole, T., Hollweg, C., Winkler, D., 2018. Trade in Global Value Chains: An Assessment of Labor Market Implications. Job Work. Pap. World Bank Gr. 1–77.

Feenstra, R.C. Hanson, G.H., 1996. Globalisation, outsourcing and wages inequlity. NBER Woking Pap. 1–13.

Fernandes, A., Kee, H.L., Winkler, D., 2020. Determinants of Global Value Chain Participation Cross-Country Evidence. World Bank Policy Res. Work. Pap. 9197, 1–54.

Fortanier, F., Miao, G., Kolk, A., Pisani, N., 2020. Accounting for firm heterogeneity in global value chains. J. Int. Bus. Stud. 51, 432–453. https://doi.org/10.1057/s41267-019-00282-0

Gao, Y. Meng, B. Suder, G. Ye, J.S.Y., 2023. Articles: Making global value chains visible: Transnational corporations versus domestically owned firms. Transnatl. Corp. 30, 1–47. https://doi.org/10.18356/2076099x-30-1-1

Gereffi, G., Fernandez-Stark, K., 2018. Global Value Chain Analysis: A Primer Second Edition1. Glob. Value Chain. Dev. Redefining Contours 21 305–342. https://doi.org/10.1017/9781108559423.012

Grossman, G., Rossi-Hansberg, E., 2008. Trading Tasks: A Simple Theory of Offshoring. Am. Econ. Rev. 98, 1978–1997. DOI: 10.1257/aer.98.5.1978

Guha-Khasnobis, B., Aditya, A., Chandna, S., 2023. Employment and global value chain participation: the Indian experience. Int. J. Econ. Policy Stud. 17, 75–94. https://doi.org/10.1007/s42495-022-00092-7

Gunnella, V., Fidora, M., Schmitz, M., 2017. The impact of global value chains on the macroeconomic analysis of the euro area. ECB Econ. Bull. 49, 75–95.

Guschanski, A., Onaran, Ö., 2023. Global Value Chain Participation and the Labour Share: Industry-level Evidence from Emerging Economies. Dev. Change 54, 31–63. https://doi.org/10.1111/dech.12749

Helpman, H., 1985. International Trade in Differentiated Middle Products, in: In Structural Adjustment in Developed Open Economies. Palgrave Macmillan, London, pp. 3–34. OI: 10.1007/978-1-349-17919-0_1

Hollweg, C., 2019. Global value chains and employment in developing economies. https://www.wto.org/english/res_e/booksp_e/gvc_dev_report_2019_e_ch3.pdf

Jones, L., Demirkaya, M., Bethmann, E., 2019. United States International Trade Commission Journal of International Commerce and Economics Global Value Chain Analysis: Concepts and Approaches. Int. Commer. Econ. 1–29.

Jones, R.W., Kierzkowski, H., 2001. Horizontal Aspects of Vertical Fragmentation, in: Cheng, L.K., Kierzkowski, H. (Ed.), Global Production and Trade in East Asia. Spinger, Boston, pp. 33–51.

Kordalska, A., Olczyk, M., Stöllinger, R., Zavarská, Z., 2022. Functional Specialisation in EU Value Chains: Methods for Identifying EU Countries' Roles in International Production Networks. wiiw Res. Reports 461, 1–51.

Kordalska, A., Olczyk, M., 2022. Upgrading low value-added activities in global value chains: a functional specialization approach. Econ. Syst. Res. 1–27. https://doi.org/10.1080/09535314.2022.2047011

Korwatanasakul, U., Baek, Y., Majoe, A., 2020. Analysis of Global Value Chain Participation and the Labour Market in Thailand: A Micro-level Analysis. ERIA Discuss. Pap. Ser. 1–21.

Kottaridi, C. Louloudi, K. Karkalakos, S., 2019. Human capital, skills and competencies: Varying effects on inward FDI in the EU context. Int. Bus. Rev. 28, 375–390. DOI: 10.1016/j.ibusrev.2018.10.008

Kucera, D., Sarna, R., 2006. Trade union rights, democracy, and exports: A gravity model approach. Rev. Int. Econ. 14, 859–882. https://doi.org/10.1111/j.1467-9396.2006.00627.x

Kummritz, V., Quast, B., 2016. Global Value Chains in Low- and Middle-Income Countries. CTEI Work. Pap. No 2016-10.

Lewandowski, P., Madoń, K., Winkler, D.E., 2023. The Role of Global Value Chains for Worker Tasks and Wage Inequality. Policy Res. Work. Pap. 10433, 1–34. https://doi.org/10.2139/ssrn.4599642

Li, J., Choi, Y., 2019. Foreign Direct Investment (FDI), GVC Participation and Trade in Value Added. Korea Trade Rev. 44, 107–125. https://doi.org/10.22659/KTRA.2019.44.5.107

Long, T.Q., Helble, M., Trang, L.T., 2019. Global Value Chains and Formal Employment in Vietnam. ERIA Discuss. Pap. Ser. ERIA-DP-20, 1–32.

Lopez-Acevedo, G., Robertson, R., 2016. Stitches to Riches?: Apparel Employment, Trade, and Economic Development in South Asia, Stitches to Riches?: Apparel Employment, Trade, and Economic Development in South Asia. The World Bank, New York. https://doi.org/10.1596/978-1-4648-0813-5

Mancini, M. Montalbano, P. Nenci, S. Vurchio, D., 2024. Positioning in Global Value Chains: World Map and Indicators, a New Dataset Available for GVC Analyses. World Bank Econ. Rev. lhae005.

Marjit, S., Mandal, B., Yang, L., 2024. New trade theory converges to the old trade theory—An elementary theoretical perspective. Int. J. Econ. Theory preprint. https://doi.org/10.1111/ijet.12412

Martínez-Galán, E. Fontoura, M.P., 2019. Global Value Chain and inward foreign direct investment in the 2000s. World Econ. 42, 175–196. https://doi.org/10.1111/twec.12660

McCaig, Brian Pavcnik, N., 2018. Export Markets and Labor Allocation in a Low-Income Country. Am. Econ. Rev. 108, 1899–1941.

Melitz, M., 2003. The Impact of Trade on Intra-industry Reallocations and Aggregate Industry Productivity. Econometrica 71, 1695–1725.

Meng, B., Wei, S.J., 2020. Measuring Smile Curves in Global Value Chains. Oxf. Bull. Econ. Stat. 82, 988–1016. https://doi.org/https://doi.org/10.1111/obes.12364

Mingyang, Y., Hankun, Y., Chen, X., Zhida, J., 2023. The impact of trade on employment: New evidence from a global value chains perspective. PLoS One 18, 1–14. https://doi.org/10.1371/journal.pone.0285681

Miningou, E.W., Tapsoba, S.J., 2020. Education Systems and Foreign Direct Investment: Does External Efficiency Matter? J. Appl. Econ. 23, 583–599.

Ndubuisi, G., Owusu, S., 2023. Global Value Chains, Job Creation, and Job Destruction Among Firms in South Africa. STEG Work. Pap. 073, 1–20.

Nikulin, D., Szymczak, S., 2020. Effect of the integration into global value chains on the employment contract in central and eastern european countries. Equilibrium. Q. J. Econ. Econ. Policy 15, 275–294. https://doi.org/10.24136/eq.2020.013

Nur, Carpa; Martínez-Zarzoso, I., 2022. The impact of global value chain participation on income inequality. Int. Econ. 169, 269–290. https://doi.org/10.1016/j.inteco.2022.02.002.

OECD, 2011. ISIC Rev. 3 Technology Intensity Definition. Mimeo.

OECD, WTO, World Bank, 2014. Global Value Chains: Challenges, opportunities and implications for policy. Rep. Prep. Submiss. to G20 Trade Minist. Meet. Sydney, Aust. 53.

Oladapo, Musibau; Raifu, I., 2022. The Effects of Global Value Chains (GVC) on Employment in Nigeria. https://dbn-bucket.s3.eu-west-2.amazonaws.com/publications/gvc-and-employment-accept-Uo5PmK63DWJMuBazM0.pdf

Pahl, S., Timmer, M.P., Gouma, R., Woltjer, P.J., 2022. Jobs and Productivity Growth in Global Value Chains: New Evidence for Twenty-five Low-and Middle-Income Countries. World Bank Econ. Rev. 36, 670–686. https://doi.org/10.1093/wber/lhac003

Pan, Z., 2020. Employment impacts of the US global value chain participation. Int. Rev. Appl. Econ. 34, 699–720. https://doi.org/10.1080/02692171.2020.1755238

Pantelopoulos, G., 2022. Higher education, gender, and foreign direct investment: Evidence from OECD countries. Ind. High. Educ. 36, 86–93.

Paul, S.; Timmer, M.; Gouma, R. ;Woltjer, P., 2019. Jobs in Global Value Chains: New Evidence for Four African Countries in International Perspective. Policy Res. Work. Pap. 8953, 1–76.

Ptas, D., 2019. Drivers of Participation. World Dev. Rep. 2020 Trading Dev. Age Glob. Value Chain. 36–63. https://doi.org/10.1596/978-1-4648-1457-0_ch2

Qiang, C.Z., Liu, Y., Steenbergen, V., 2021. An Investment Perspective on Global Value Chains, An Investment Perspective on Global Value Chains. World Bank Group, Washington. https://doi.org/10.1596/978-1-4648-1683-3

Sahiti, A. Ahmeti, S. Ismajli, H., 2018. A review of empirical studies on FDI determinants. Balt. J. Real Estate Econ. Constr. Manag. 6, 37–47.

Shepherd, B., Stone, S., 2013. Global Production Networks and Employment: A Developing Country Global Production Networks and Employment. OECD Trade Policy Pap. 154, 1–31.

Shepherd, B., 2013. Global Value Chains and Developing Country Employment: A Litterature Review. OECD Trade Policy Pap. 4–20.

Shingal, A., 2015. Labour market effects of integration into GVCs: Review of literature. R4D Work. Pap. 10, 1–23.

Stollinger, R., 2019. Testing the Smile Curve: Functional Specialisation in GVCs and Value Creation. WIIW Work. Pap. 163, 1–45.

Szymczak, S., 2024. The impact of global value chains on wages, employment, and productivity: a survey of theoretical approaches. J. Labour Mark. Res. 58. https://doi.org/10.1186/s12651-024-00367-w

Szymczak, S., Wolszczak-Derlacz, J., 2021. Global value chains and labor markets– simultaneous analysis of wages and employment. Econ. Syst. Res. 34, 69–96. https://doi.org/10.1080/09535314.2021.1982678

Taglioni, D., Winkler, D., 2016. Making Global Value Chains Work for Development. Trade and Development. World Bank, Washington.

Villaverde, J. Maza, A., 2015. The determinants of inward foreign direct investment: Evidence from the European regions. Int. Bus. Rev. 24, 209–223.

Winkler, D., Kruse, H., Aguilar Luna, L., Maliszewska, M., 2023. Linking Trade to Jobs, Incomes, and Activities: New Stylized Facts for Low- and Middle-Income Countries. Policy Res. Work. Pap. 53.

Wolszczak-Derlacz, J. Nikulin, D. Szymczak, S., 2023. Global value chains and wages under different wage setting mechanisms. Compet. Chang. 27, 809–829. https://doi.org/10.1177/10245294221131942

Wooldridge, J., 2002. Econometric Analysis of Cross Section and Panel Data. MIT Press, Cambridge.

Zhao, J., 2021. Impacts of global value chains on foreign direct investment (The case of Asian developing countries). Econ. Bull. 41, 1139–1152.

## Annex

### Table A 1 – Variables definitions and sources

| Variable | Definition | Source |
|---|---|---|
| Log of number of jobs created by FDIs | Number of jobs created by single FDI inflow project, aggregated at the NACE Rev.2 two-digit level | FDI Markets Crossborder Monitor database |
| GVC participation forward | Domestic value added in gross exports divided by Value added | TiVA database, own calculation (winsorized) |
| GVC participation backward | Gross imports of intermediate products divided by Gross output | TiVA database, own calculation (winsorized) |
| GVC position | Upstreamness of a country/sector - the distance of its productions from the final demand; divided by Downstreamness - the distance from the factors of production (sources of value-added) | TiVA database, calculation by Mancini et al. (2024) |
| GDP growth rate | Growth rate of GDP expressed in %, by country | World Development Indicators |
| Log of GDP per person employed | GDP (international dollars, PPP) divided by the number of employed, by country, logged | World Development Indicators |
| Trade | Trade (exports plus imports) divided by GDP | World Development Indicators |
| Government spending on education | Government expenditure on education divided by GDP | World Development Indicators |
| Log of sectoral productivity | Value added divided by Number of employees, by sector (NACE Rev.2, two-digit), logged | UNIDO Industrial Database, own calculation |
| Log of sectoral average wage | Average wage and supplements paid out to employees, by sector, logged | UNIDO Industrial Database, own calculation |

*Source: Authors' presentation.*

### Table A 2 – Descriptive statistics

| Variable | Obs. | Mean | Std.Dev. | Min | Max |
|---|---|---|---|---|---|
| Log of number of jobs created by FDI | 6,022 | 5.13 | 1.73 | 0.00 | 11.86 |
| GVC participation forward | 5,897 | 0.67 | 0.42 | 0.00 | 1.87 |
| GVC participation backward | 5,902 | 0.30 | 0.36 | 0.00 | 4.46 |
| GVC position | 6,025 | 1.00 | 0.26 | 0.36 | 2.38 |
| GDP growth rate (%) | 6,025 | 2.08 | 3.76 | (14.84) | 24.48 |
| Log of GDP per person employed | 6,025 | 11.47 | 0.34 | 10.62 | 12.68 |
| Trade (% of GDP) | 6,025 | 108.66 | 49.31 | 45.42 | 382.35 |
| Government spending on education (% of GDP) | 5,917 | 5.00 | 1.09 | 2.32 | 8.56 |
| Log of sectoral productivity (sectoral value added per employee) | 4,625 | 11.45 | 1.57 | 1.23 | 17.78 |
| Log of sectoral average wage | 4,657 | 21.22 | 2.19 | 11.51 | 27.19 |

*Source: Authors' presentation.*

Table A 3 – Correlogram of variables

| | Log of number of jobs created by FDI | GVC participation forward | GVC participation backward | GVC position | GDP growth rate (%) | Log of GDP per person employed | Trade (% of GDP) | Government spending on education (% of GDP) | Log of sectoral productivity | Log of sectoral average wage |
|---|---|---|---|---|---|---|---|---|---|---|
| Log of number of jobs created by FDI | 1.000 | | | | | | | | | |
| GVC participation forward | 0.229 | 1.000 | | | | | | | | |
| GVC participation backward | (0.058) | 0.024 | 1.000 | | | | | | | |
| GVC position | (0.065) | (0.012) | (0.070) | 1.000 | | | | | | |
| GDP growth rate (%) | 0.103 | (0.064) | 0.069 | 0.020 | 1.000 | | | | | |
| Log of GDP per person employed | (0.222) | 0.074 | (0.085) | 0.084 | (0.269) | 1.000 | | | | |
| Trade (% of GDP) | (0.258) | 0.050 | (0.028) | 0.104 | (0.203) | 0.441 | 1.000 | | | |
| Government spending on education (% of GDP) | (0.066) | 0.215 | 0.173 | 0.021 | 0.185 | 0.071 | (0.017) | 1.000 | | |
| Log of sectoral productivity | 0.078 | 0.159 | (0.002) | (0.020) | (0.087) | 0.208 | 0.291 | 0.174 | 1.000 | |
| Log of sectoral average wage | 0.2875 | 0.1736 | -0.2857 | -0.0716 | -0.1574 | 0.2382 | 0.1382 | -0.1833 | 0.7489 | 1.000 |

Source: Authors' presentation.